\documentclass[twocolumn,superscriptaddress,showpacs,preprintnumbers,amsmath,amssymb]{revtex4}
\usepackage{graphicx}
\usepackage{dcolumn}
\usepackage{bm}

\begin{document}

\title{ Correlation Effects and Structural Dynamics in the $\beta$-Pyrochlore
 Superconductor KOs$_{2}$O$_{6}$}
\author{J. Kune\v{s}}
\affiliation{Department of Physics, University of California, One Shields Avenue,
Davis CA 95616, USA}
\affiliation{Institute of Physics,
Academy of Sciences of the Czech Republic, Cukrovarnick\'a 10,
162 53 Praha 6, Czech Republic}
\author{T. Jeong}
\affiliation{Department of Physics, University of California, One Shields Avenue,
Davis CA 95616, USA}
\author{W.\,E. Pickett}
\affiliation{Department of Physics, University of California, One Shields Avenue,
Davis CA 95616, USA}
\date{\today}

\date{\today}

\begin{abstract}
Electronic, magnetic, and dynamical properties of the new superconducting
$\beta$-pyrochlore KOs$_2$O$_6$ and related RbOs$_2$O$_6$ and CsOs$_2$O$_6$
compounds are calculated and compared with experiment 
and contrasted with structurally related spinel pyrochlores.  The calculated
susceptibility Stoner enhancement (110\%) and thermal mass enhancement 
$\lambda$ = 2.5-3 reflect moderate but perhaps important Coulomb
correlations.  The K$^{+}$ ion optic mode is found to be unstable, allowing
large excursions of 0.5-0.6~\AA~from its ideal site of the K ion along $\langle 111 \rangle$
directions. This dynamical
mode is much less anharmonic in the isostructural Rb and Cs compounds
(with larger cations), perhaps
accounting for their progressively lower values of T$_c$.
Electron scattering from this very anharmonic mode may be the cause of the
anomalous concave-downward resistivity that is seen only in KOs$_2$O$_6$.  
\end{abstract}


\maketitle

\section{Introduction}
Transition metal oxides with metal ions (T) lying on a
pyrochlore sublattice display 
a wide variety of behavior. In the 1970s LiTi$_{2}$O$_{4}$ was the 
anomalous ``high $T_c$'' oxide superconductor amongst 
intermetallics, with its $T_c=13K$. 
The vanadate LiV$_{2}$O$_{4}$ became the first (and still essentially
the only) true heavy fermion 
metal based on $d$ electrons rather than $f$ electrons. Considerable 
theoretical speculation on the microscopic basis for this 
heavy fermion behavior has left no consensus. Other members of the 
pyrochlore structure, such as CuIr$_{2}$S$_{4}$ and Tl$_{2}$Ru$_{2}$O$_{7}$,
exhibit charge ordering and accompanying structural adjustment, possibly
associated with the high symmetry of the 
undistorted structure. The ``pyrochlore'' sublattice occupied by the 
transition metal ions is comprised of corner-sharing tetrahedra that are
known to be highly frustrating for nearest neighbor antiferromagnetic (AFM)
spin couplings.\cite{Ramirez,Canals}

Recently Yonezawa {\it et al.}\cite{Yonezawa,Hiroi} reported the discovery
of a superconductor KOs$_{2}$O$_{6}$ with a $T_{c}= 9.6$K,
with $\beta$-pyrochlore structure, a new variant of the pyrochlore class,
and superconductivity was quickly obtained in isostructural and
isoelectronic RbOs$_2$O$_6$
(6.3 K)\cite{rbos2o6} and CsOs$_2$O$_6$ (3.2 K)\cite{csos2o6} as well, suggesting
new physics generic to this structural variant.  
The other example of superconductivity in $4d/5d$ pyrochlore oxides is 
Cd$_{2}$Re$_{2}$O$_{7}$ with $T_{c}=1$ K.\cite{Hanawa}
There is no obvious indication of strong enhancement by correlation effects
in the quasiparticle mass for the $\beta$-pyrochlore compounds. 
Using the specific heat jump $\Delta C$ at T$_c$
and the weak coupling relation $\Delta C/T_c = 1.43 \gamma$, Hiroi {\it et al.}
\cite{Hiroi} obtained a linear specific heat coefficient $\gamma$ = 19
mJ/K$^2$ mol-Os, which is not especially large. The magnitude has been confirmed
for RbOs$_2$O$_6$ by Br\"uhwiler {\it et al.} \cite{bru04} who obtained $\gamma$=
17 mJ/K$^2$ mol-Rb from the heat capacity. However, one highly
unusual feature is that,
while the resistivity $\rho(T)$ of both RbOs$_2$O$_6$ (Ref. \cite{rbos2o6})
and CsOs$_2$O$_6$ (Ref. \cite{csos2o6})
have the normal upward curvature at low
temperature, KOs$_2$O$_6$ with its higher $T_c$ has
a very peculiar concave downward shape\cite{Hiroi} of $\rho(T)$ immediately above $T_c$
extending to 200 K.

Conventional pyrochlore oxides have 
the chemical formula A$_{2}$T$_{2}$O$_{7}$ or more descriptively
A$_{2}$T$_{2}$O$_{6}$O$'$, 
where A is a larger cation, T is a smaller transition metal
cation, and O$^{\prime}$ is an oxygen
ion in a large tetrahedral site rather than forming the octahedron surrounding
the transition metal ion.  The distinguishing feature is the T pyrochlore sublattice,
a network of corner-sharing tetrahedra that has been widely studied in the
context of frustrated antiferromagnetism. 
Yonezawa {\it et al.}\cite{Yonezawa} 
produced the $\beta$ variant with the general formula AT$_{2}$O$_{6}$
where A is a large monovalent alkaline metal cation.
KOs$_{2}$O$_{6}$ has the same 
space group $Fd\bar{3}m$ as conventional pyrochlore, 
but the large cations K$^{+}$, Rb$^{+}$, Cs$^{+}$ 
are located (unexpectedly) at the O$^{\prime}$ 
site of the conventional pyrochlore structure, thus becoming the 
$\Box_2$Os$_2$O$_6$K variant of conventional pyrochlore (where $\Box$ denotes
an empty A site).  
KOs$_{2}$O$_{6}$ has the lattice constant 10.101~\AA~and two formula units per
fcc primitive cell.

\begin{figure}
\includegraphics[height=8.5cm,angle=0]{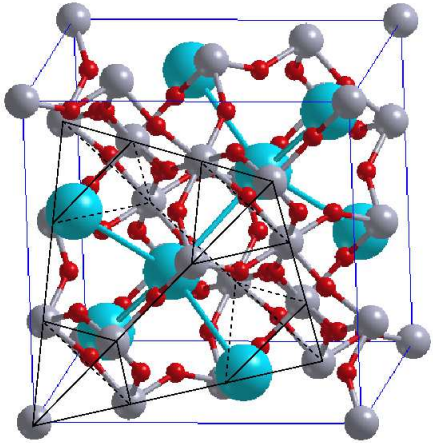}
\caption{\label{fig:str}(color online) 
The crystal structure of the $\beta$-pyrochlore superconductor KOs$_{2}$O$_{6}$.
The largest atoms are K, the smallest dark spheres are O and are connected 
by sticks with the midsize atoms, Os. The pyrochlore sublattice formed by Os tetrahedra and
truncated-tetrahedral cavities containing K ions is highlighted.} 
\end{figure}

Although the T ions lie on the same pyrochlore sublattice as in the 
closely related spinel system AT$_2$O$_4$, and are the 
centers of TO$_6$ octahedra, there are {\it qualitative} differences:
in the spinel structure all the {\it edge-sharing} TO$_6$ octahedra are aligned 
along the cubic axes with T atoms being bridged by two O ions, while 
in the conventional pyrochlore and its $\beta$ variant there are four different
orientations of the {\it vertex-sharing} TO$_6$ octahedra and each pair
of T ions is bridged by a single O ion. The four different TO$_6$ octahedra are
rhombohedrally distorted along the one of the four cubic body diagonals by an 
amount determined by the O internal parameter (u=0.3125 corresponds to an 
equilateral octahedron).
The pyrochlore lattice of the T sites can be viewed as a 3D generalization of
the 2D Kagome lattice (see Fig. \ref{fig:str}).

In this paper we initiate the theoretical study of the KOs$_{2}$O$_{6}$ class
of compounds,
focusing on the electronic structure in comparison to the
well studied conventional pyrochlore and spinel systems.  The tight-binding
description differs in important ways from the spinel (LiV$_2$O$_4$, say)
counterpart, and the density of states N(E$_F$) at the Fermi level
E$_F$ is not high. In fact it is below the average for the $t_{2g}$ band, due to
E$_F$ lying in a valley, suggesting either electron or hole doping could enhance
T$_c$ by raising N(E$_F$).  We further
describe the discovery of an extreme dynamical
instability corresponding to displacements of 0.5~\AA\ or more of the alkali ion
along $\langle 111 \rangle$ direction.

\section{Method}
We have used both the full-potential 
nonorthogonal local-orbital minimum-basis (FPLO) scheme within the local 
density approximation (LDA)\cite{Koepernik} and the full potential linearized
augmented plane wave (LAPW) plus local orbitals method as implemented in 
Wien2k.\cite{wien2k} The exchange and correlation potential of Perdew and Wang\cite{perdew}
was used.  
In FPLO,
K $3s,3p,4s,4p,3d$ states, Os  $4s,4p,4d,4f,5s,5p, 6s,6p$ and 
O $2s,2p,3d$ were included as 
valence states. 
The inclusion of the relatively extended $3s,3p,4s,4p,3d$ semicore 
states as band states was done because of the considerable overlap of these 
states on nearest neighbors. 
The LAPW basis set is characterized by the atomic sphere radii of 2.05 bohr for 
K and Os and 1.4 bohr for O, plane-wave cut-off R$_{mt}$K$_{max}$=5.5 and
K-$s$, -$p$, Os-$s$, -$p$, -$f$ and, O-$s$ local orbitals. 
The self-consistent potentials were calculated using
256 k points in the irreducible zone. The K, Os and O atoms are located at positions
$(\frac{3}{8},\frac{3}{8},\frac{3}{8}), (0,0,0),$and
$(u,\frac{1}{8},\frac{1}{8})$ with site symmetries $\bar{4}$3m, $\bar{3}$m, and mm2 respectively.
We obtained the internal coordinate $u=0.317$ by minimizing the total energy
with FLAPW method. The corresponding deviation of the O-Os-O bond angle from the 
right angle is $\pm$ 1.8$^o$. Similar $u$ values of 0.316 and 0.315 were obtained for
RbOs$_2$O$_6$ and CsOs$_2$O$_6$ respectively, corresponding to O-Os-O bond angle deviations 
for right angle of 1.6$^o$ and 1.4$^o$. Due to the relaxation of the O-Os-O bond angle the relative
differences of Os-O bond lengths are less the relative differences of the lattice constants for
the three systems.

\section{Results and Discussion}
\subsection{Band and Density of States}
\begin{figure}
\includegraphics[height=9cm,angle=-90]{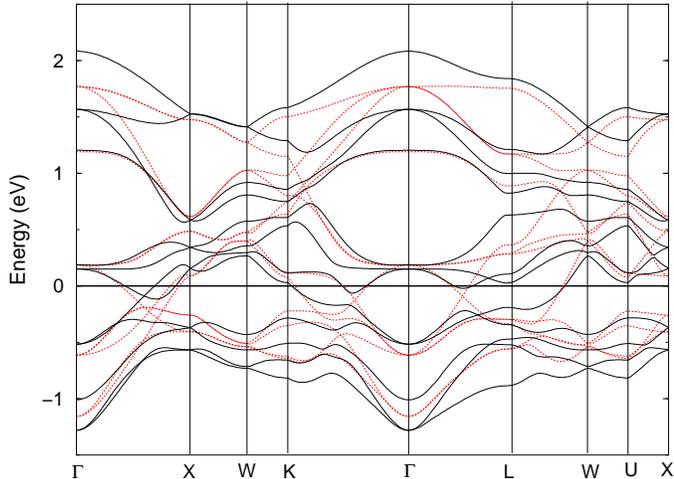}
\caption{\label{fig:band} The Os $t_{2g}$ band complex of the superconductor KOs$_2$O$_6$ along
fcc symmetry lines.  The formal $d^{2.5}$ configuration results in 5/12 
filling of these twelve bands. The relativistic bands are marked with full (black)
lines, the scalar-relativistic with dotted (red) line.} 
\end{figure}

\begin{figure}
\includegraphics[width=8cm]{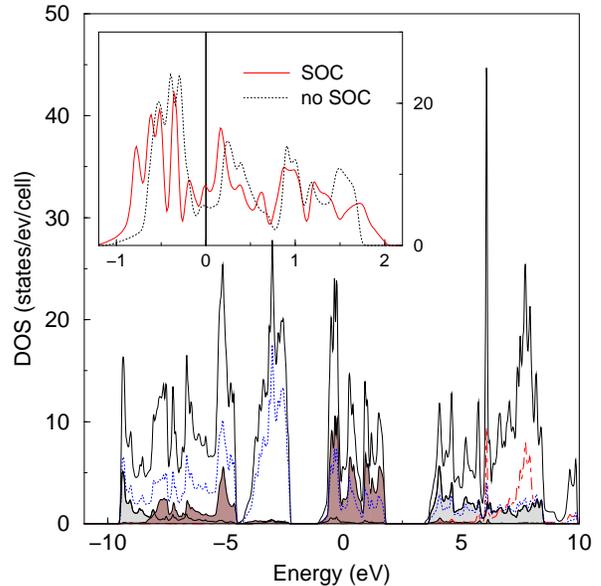}
\caption{\label{fig:dos}(color online) Scalar-relativistic density of states (solid line) 
and the site and orbital projected densities
of states of KOs$_2$O$_6$: Os-$t_{2g}$ (dark shading), Os-$e_g$ (light shading),
O (dotted line), and K (dot-dashed line). In the inset we show in detail the effect of 
spin-orbit coupling to the density of states of the $t_{2g}$ manifold.} 
\end{figure}

The guideline K$^{+}$Os$^{p+}_2$O$^{2-}_6$ leads to the
formal valence $p=5.5$, or a $d^{2.5}$ occupation (out of the six $t_{2g}$
states, including spin). As we see in Fig. \ref{fig:band}, this puts 
E$_F$ within the $t_{2g}$ bands, somewhat below half-filling of the 
threefold orbitally degenerate complex of bands.  
Surrounding E$_F$ we have the $t_{2g}$ manifold (as in both pyrochlore A$_2$T$_2$O$_7$
and spinel AT$_2$O$_4$ systems) of width 3.2 eV (2.9 eV), and above it
and separated by 1.6 eV (1.8 eV) is the $e_{g}$ manifold. 
The ``O $2p$ bands'' (with both $e_g$ and $t_{2g}$ character mixed in)
are fully occupied and separated from the $t_{2g}$ bands by 0.7 eV (1.5 eV) as shown
in Fig. \ref{fig:dos} (scalar-relativistic values are shown in the brackets). 
The non-cubic part of the $\bar{3}$m crystal field on the Os site 
is weak and the band structure 
shows no apparent further splitting of the $t_{2g}$ complex.
We show a blowup of the $t_{2g}$ bands around $E_{F}$ of
KOs$_{2}$O$_{6}$ in Fig. \ref{fig:band}.
The twelve bands (3 $t_{2g}$ $\times$ 4 Os ions) are 5/12 filled.
For comparison we have also calculated the band structures of the other $\beta$-pyrochlore 
superconductors, CsOs$_{2}$O$_{6}$ and RbOs$_{2}$O$_{6}$. The band structures are 
extremely similar, with the same bandwidth, differing only in fine details. 

In Fig. \ref{fig:dos} we display the density of states (DOS) of KOs$_{2}$O$_{6}$.
The calculated Fermi level $E_F$ lies within a valley, giving a value of
N(E$_F$)=8.3 states/eV/unit cell (scalar-relativistic value) that is slightly above the average value
of 7.5 states/eV in the $t_{2g}$ bands, and $E_F$ lies 0.3 eV above the peak in N(E) arising
from the rather flat bands at the bottom of the complex.
The DOS plot also indicates that the
K states appear nowhere close to the Fermi level and do not exhibit
mixing with the Os $d$ states, which limits the role of the alkali
ion to a donor of an electron and a source of electrostatic potential
(and thus a possible scatterer).

\subsection{Tight Binding Representation}
In order to understand the band structure of the $t_{2g}$ complex, 
we are thus left to consider the Os lattice bridged by O ions, suggesting 
a simple tight-binding description. Crystal field
splitting reduces the number of relevant orbitals to three $t_{2g}$ states per Os site. 
Separation of the O bands suggests that the O $p$ states can
be ``integrated out'' and taken into consideration implicitly
through the effective Os-Os hopping amplitude. 
Fujimoto\cite{flatbands} has shown that a single ($s$) state on a pyrochlore 
lattice with near neighbor coupling only leads to two
dispersing bands and two absolutely flat bands, which reflects the close relation
to the Kagome lattice (which has a single flat band). 
Singh {\it et al.}\cite{djs} considered a much more realistic $dd\sigma, dd\pi$ 12-band
model for LiV$_2$O$_4$, and noted that for $dd\pi$ =$\frac{3}{8} dd\sigma$, two of the twelve
bands become flat. Unlike for the spinel structure,
in the $\beta$-pyrochlore structure the principle axes of the $t_{2g}$ 
states are not aligned with the cubic
axes so such a model does not apply directly to KOs$_{2}$O$_{6}$.

\begin{figure}
\includegraphics[height=8cm,angle=270]{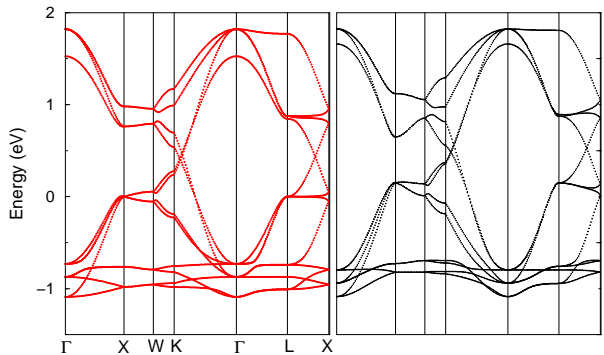}
\caption{\label{fig:model} Tight-binding bandstructure: $t_{2g}$-to-$t_{2g}$
hopping via oxygen only (left panel), including $e_g$-to-$e_g$ and
$e_g$-to-$t_{2g}$ hopping (right panel).}
\end{figure}

Based on the geometry of the Os-O-Os bonds we have developed a tight-binding model
with three $t_{2g}$ states on four Os sites in the unit cell.
In an equilateral octahedron only $dp\pi$ hopping is possible for $t_{2g}$ orbitals
(e.g. for O atom at (001) vertex of the local octahedron the $d_{xz}\rightarrow p_x$ and
$d_{yz}\rightarrow p_y$ hopping is allowed by symmetry).
The effective hopping amplitude between $t_{2g}$ orbitals
on neighboring Os sites is then determined  by this single parameter and the relative
orientation of the corresponding OsO$_6$ octahedra.
This model (see Fig. \ref{fig:model}) provides
a reasonable band picture except that (1) the
order of the T$_{2g}$ and T$_{1g}$ triplets (second and third levels from the
bottom of the $t_{2g}$ complex) at the zone center are
interchanged and thus the band topology in this region (at and below E$_F$) is
incorrect, (2) the four lower bands are too flat, and (3) unphysical degeneracies
occur at the zone boundary points X and L (this last point is minor).
Inclusion of the coupling to $e_g$ states readily corrects the ordering of the
T$_{1g}$ and T$_{2g}$ triplets and thus the band topology as shown in 
Fig. \ref{fig:model} (right panel). 
Explicit incorporation of the O 2$p$ states into a total 56 band
model gives no appreciable change of the Os derived bands, indicating that the O $2p$ bands can
be integrated out fairly accurately. The additional dispersion of the flat
bands is likely to be due to direct Os-Os hopping. A rough estimate of this effect 
can be obtained from the Os lattice without O, yielding the $d$ bandwidth less then 1 eV. 

\subsection{Fermiology}
An appreciable effect of spin-orbit coupling is to be expected in Os $5d$ bands.
While the relativistic bands are qualitatively similar to the scalar-relativistic
ones, there are some important differences. Removal of some band crossings
leads to flatter bands and additional peaks in the density of states.  
As a result the density of states at the Fermi level is enhanced (see below) and
the band structure in the vicinity of Fermi level becomes more sensitive
to the oxygen position as we discuss below.

\begin{figure}
\includegraphics[width=8cm]{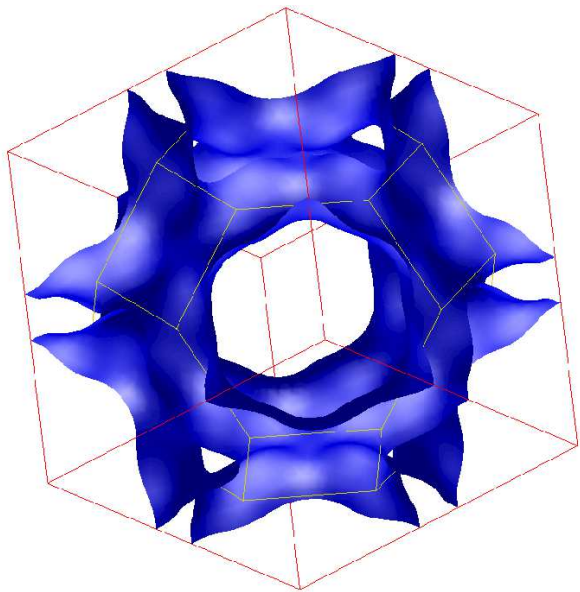}
\caption{\label{fig:fermi1} The connected the Fermi surface with necks 
along the three-fold axis (spin-orbit coupling included).}
\end{figure}

\begin{figure}
\includegraphics[width=8cm]{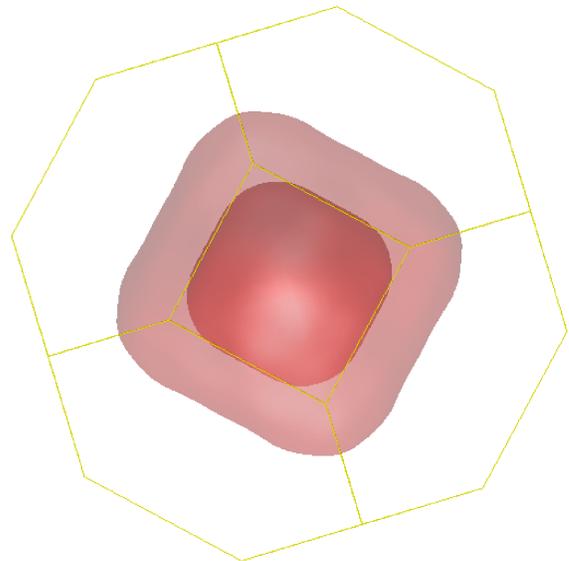}
\caption{\label{fig:fermi2} The closed sheets of the Fermi surface centered
on the $\Gamma$ point (spin-orbit coupling included).}
\end{figure}

\begin{figure}
\includegraphics[height=8cm,angle=270]{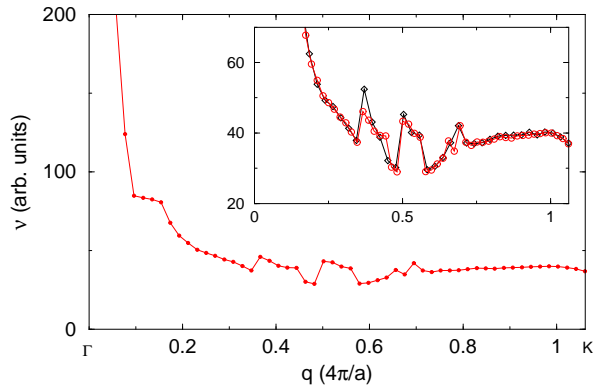}
\caption{\label{fig:nesting} The low frequency limit of the generalized susceptibility
$\nu({\mathbf q})$ calculated along the $\Gamma$-K line obtained with 440$\times$440$\times$440
k-points (circles) in the cubic Brillouin zone. In the inset we compare with a curve obtained
with 160$\times$160$\times$160 k-mesh (diamonds).}
\end{figure}

There are two bands crossing the Fermi level in Fig. \ref{fig:band}. The first band gives rise
to two closed sheets of the Fermi surface centered at the $\Gamma$ point, which
enclose the electron part of the reciprocal space in between of them (see Fig. \ref{fig:fermi1}).
The second band leads to a connected Fermi surface with necks along the
three-fold axis (see Fig. \ref{fig:fermi2}). Visual inspection of the
closed sheets of the Fermi surface suggests a possibility of partial nesting,
in particular along the $\Gamma-K$ direction. In order to pursue this possibility
we have calculated the imaginary part of generalized susceptibility in the
low energy limit
\begin{align}
\label{eq:lim}
\operatorname{Im}\chi({\mathbf q},\omega)&=\pi\omega \sum_{\mathbf k}\delta(\epsilon({\mathbf k})-
E_F)\delta(\epsilon({\mathbf k+\mathbf q})-E_F)\\
&=\pi\omega\nu({\mathbf q}).
\end{align}
We emphasize that a large number of k-points is required to get reliable results,
see the caption to Fig. \ref{fig:nesting}.
The calculated Im$\chi({\mathbf q})$ along $\Gamma$-K has some sharp Fermi surface
related structure, but the peaks are only $\pm$15\% from a smooth background and so do
not suggest electronic instabilities.

For the three $\beta$-pyrochlores reported so far (K, Rb, Cs) \cite{Yonezawa,
rbos2o6,csos2o6} the transition temperature $T_c$ and the lattice parameter
vary inversely as shown in Table \ref{tab:tc}.
It is plausible to assume that the structural change is due to the size effect of 
alkaline metal ions, because the bands near the Fermi level consist of 
Os 5d orbitals with minor contribution from O 2p orbitals.
The decrease of $T_{c}$ with 
increasing volume under negative chemical pressure 
is in contrast to the case of conventional BCS superconductivity in a 
single band model, where the $T_{c}$ increases under negative pressure, 
because the density of states increases (see Table \ref{tab:tc}).  
\begin{table}[b]
\caption{\label{tab:tc} The experimental lattice constants and transition
temperatures \cite{Yonezawa,
rbos2o6,csos2o6} together with the calculated densities of states at the
Fermi level N(E$_F$) normalized per Os atom and the corresponding
unrenormalized linear specific heat coefficient $\gamma_b$ per mole Os.}
\begin{tabular*}{\columnwidth}{|l|c|c|c|c|}
\hline
Compound & a(\AA) & T$_c$(K) & N(E$_F$)(Ry$^{-1}$) & $\gamma_b$(mJ/K$^2$)\\
\hline \hline
KOs$_2$O$_6$ & 10.101 & 9.6 & 28.2 & 4.9 \\
RbOs$_2$O$_6$ & 10.114 & 6.3 & 28.6 & 5.0 \\
CsOs$_2$O$_6$ & 10.149 & 3.2 & 30.3 & 5.3 \\
\hline
\end{tabular*}
\end{table}

Hiroi {\it et al.}\cite{Hiroi} have inferred the linear specific heat
coefficient for KOs$_2$O$_6$ of $\gamma$=19 mJ/K$^{2}$ mole-Os from 
the heat capacity jump at $T_{c}$ 
assuming the weak coupling relation $\Delta C/T_{c} = 1.43 \gamma$. 
A similar value has recently been obtained by Br\"uhwiler {\it et al.} \cite{bru04}
for RbOs$_2$O$_6$ $\gamma$=17 mJ/K$^{2}$ mole-Os directly from the
heat capacity. 
The calculated value of $N(E_{F})$=28.2 states/Ry/Os for KOs$_2$O$_6$
corresponds to
a bare value $\gamma_b$=4.9 mJ/K$^{2}$ mole-Os. The corresponding thermal 
mass enhancement $\lambda$ given by $\gamma / \gamma_b=1+\lambda$ 
gives a large renormalization $\lambda =2.9$. The value $\lambda =2.4$ is obtained
from the data for RbOs$_2$O$_6$. 
This renormalization 
includes the phonon contribution but most likely is due primarily to 
electronic processes. 

Important insight into magnetic fluctuation effects is provided by 
the enhancement of the bare Pauli susceptibility in a metal. We have
evaluated, within the density functional theory formalism, the Stoner 
enhancement of the susceptibility 
$\chi=\frac{\chi_{\circ}}{(1-IN(E_{F}))}\equiv S\chi_{\circ} $,
where $\chi_{\circ}=2\mu_{B}^{2}N(E_{F})$ 
is the non-interacting susceptibility 
and $S$ gives the electron-electron enhancement in terms of the Stoner 
constant $I$. We have calculated $I$ using both the Janak-Vosko-Perdew theory \cite{jan77}
and fixed spin moment calculations, obtaining a value of $S=2.15\pm 0.05$, which
does not indicate any ferromagnetic instability of the paramagnetic groundstate. 
The spin-orbit coupling was neglected in this calculation.

\subsection{Dynamical Instability}
\begin{figure}
\includegraphics[height=7cm,angle=270]{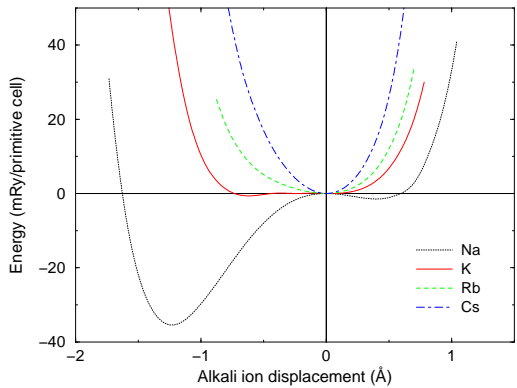}
\caption{\label{fig:amode}(color online) Energy surface for a displacement $\Delta$ of alkali ions along
$\langle 111 \rangle$ direction. The
curves were obtained by spline interpolation of energies and forces evaluated
at approximately ten points. Note the very large ($\approx$ 0.5 eV) instability
of the Na ion.}
\end{figure}

In light of interesting but perhaps limited role of correlation effects
(nothing like heavy fermion behavior), we have 
begun to pursue the character of electron-phonon coupling as a pairing
mechanism. Since K$^+$ is a bare charge in a substantial hole in the $\beta$-pyrochlore
lattice, we have calculated the energy surface and deformation potential for
a K-K ``bond stretching'' motion of the K ions, which lie on a diamond
sublattice (of course, there is no K-K bond, the ions being ionized and also separated by 
$d_{K-K} = \frac{\sqrt{3}}{4}a$ = 4.33~\AA). The result we find is a dynamical 
instability of the K ion.  Although the force vanishes by symmetry for the
ideal structure, for increasing separation of K ions lying along the $\langle 111 \rangle$
direction the energy decreases. 
The energy is minimized only after a displacement
of the K ion by 0.65~\AA!  This motion is directed along $\langle 111 \rangle$ channels in the
Os$_2$O$_6$ system, and this crystal structure may not be stable for the smaller
alkali cations Na and Li simply because they do not stay put near the 
ideal site. In Fig. \ref{fig:amode} we show the energy as a function of the
alkali ion displacement in this mode for K, Rb and Ce as 
well as fictitious Na compounds (the lattice
constant of KOs$_2$O$_6$ was used for the Na compound). Note
that the curves are upper bounds, since allowing the Os and O atoms to relax
at any displacement would only lower the energy.

The high symmetry position is found to be unstable for Na, while a very flat
energy surface is found for K over a large range of displacements. The Rb system
exhibits significant anharmonicity, which is reduced when going to Cs.
Large differences in the energy surface between systems which have 
very similar lattice constants and band structures can be understood as follows.
If only the effect of the electrostatic potential on the nucleus
at the alkali metal site was considered (i.e. if the electron charge was frozen)
the site would be dynamically unstable although the force vanishes.  
This is reflected in the fact 
that the first non-spherical term in the site expansion of the electrostatic
potential is a cubic polynomial, indicating an inflection point.
The site is thus stabilized due to electronic relaxation which
is accomplished by mixing of the outer alkali ions orbitals
with the orbitals on its neighbors.  
This explains the pronounced difference between Na, for which the $2p$ orbitals
are substantially more localized than the $5p$ orbitals of Cs.
Note that shape of the instability or anharmonicity corresponds with the
effect of the electrostatic repulsion of the four neighboring alkali ions 
(which form a tetrahedron), which tends to move the atom in the center 
away from the vertices.

In addition to the alkali ion displacement we have investigated the symmetric O mode, which corresponds
to varying the internal parameter around its equilibrium value of 0.317. This Raman-active mode
correspond to a simultaneous rhombohedral distortion of the OsO$_6$ octahedra along a cubic 
body diagonal (different diagonal for each of the four Os atoms in the unit cell). Locally
the O atom is moving perpendicular to the line connecting its nearest-neighbor Os pair. The
calculated frequency is 65 meV=525 cm$^{-1}$. Similar frequencies of 65 meV and 64 meV were obtained
for RbOs$_2$O$_6$ and CsOs$_2$O$_6$ respectively.
The deformation potential for the band
crossing the Fermi level calculated at L point (very near the Fermi level) 
amounts to $\Delta\epsilon_k/\Delta R$=1.8 eV\AA$^{-1}$,
where $\Delta R$ is the displacement of each O ion. 
The presence of two bands in the vicinity of the Fermi level (at the L point and close to
the center of $\Gamma$-L line) makes the Fermi surface rather sensitive to this
oxygen mode. These two bands move in mutually opposite directions as the oxygen is moved
away from equilibrium.
As a consequence displacement of the O atom by less then 0.05 \AA~results in appearance of an additional
electron pocket centered at the L point followed by 
sticking together and opening of holes along $\Gamma$-L direction in the sheets centered
at $\Gamma$ point.

\section{Discussion and Summary}
In this paper we have analyzed the electronic structure of the
$\beta$-pyrochlore KOs$_2$O$_6$, which is nearly identical to those
of the Rb and Cs compounds.  The Os $t_{2g}$ states are well separated from
the more tightly bound O 2p states, and also separated from the
unoccupied $e_g$ states, leaving the focus on the complex of twelve
$t_{2g}$-derived bands.  Spin-orbit coupling has a large effect on
N(E$_F$), increasing it by 60\% over its value if S-O coupling is
neglected (other regions will have decreased values of N(E)).
This system has often been compared to the other pyrochlore structure
superconductor Cd$_2$Re$_2$O$_7$, which has T$_c$ = 1 K.
The $t_{2g}$ bands of the two compounds have very much the same shape,
with those of KOs$_2$O$_6$ being 20\% wider \cite{sin02}.  The different band filling
of these compounds precludes serious comparison of their superconducting
behaviors, and the more interesting comparison is the factor of three
variation in T$_c$ within the $\beta$-pyrochlore class when they all have
very similar electronic structures.
                                                                                                           
A single parameter tight binding model provides a starting point
for an effective model Hamiltonian, but leaves the lower four bands
flatter than observed.  This aspect, as well as ordering of bands
at k=0, can be improved by including coupling between the t$_{2g}$
and $e_g$ states.  The calculated Fermi surfaces consist of
two closed surfaces at the zone center, with flat portions, and
a large surface that is open along the $\langle 111 \rangle$ directions.  The
closed surfaces give rise to nesting features at specific values
of q along the $\langle 110 \rangle$ direction, but while sharp the features are
not very strong.
                                                                                                           
Investigation of the strength of electronic correlations has been
initiated here.  Comparison of the measured linear specific heat
coefficient $\gamma$ for RbOs$_2$O$_6$ with the calculated value of N(E$_F$)
leads to a dynamic quasiparticle mass enhancement $\lambda$ = 2.4,
indicating important correlation effects but far from heavy fermion
type of behavior.  Calculation of the Stoner enhancement of the
q=0 susceptibility gives a factor of two, again indicating
important correlation corrections but nothing close to a ferromagnetic
instability.
                                                                                                           
The only substantial difference between the K, Rb, and Cs compounds
that we have found is the degree of (in)stability of the alkali cation
in its tetrahedral interstitial site. The potential for K ion motion
along the $\langle 111 \rangle$ directions is flat over a distance of almost 1~\AA,
indicating an extremely floppy ion whose motion may provide
the scattering of carriers that is reflected in peculiar
concave-downward resistivity at low temperature.  The Rb and Cs ions
are progressively more stable, and their resistivity behavior is
more conventional.  Calculation for the Na compound (not yet
reported) indicates a seriously unstable potential surface, with
the minimum being more than 1~\AA~away from the normal high symmetry
site of the cation.  This instability may make the Na compounds
unstable (and unsynthesizable) in this structure.

The close similarity of the electronic structures of the compounds in
the KOs$_2$O$_6$ series, as well as almost identical 
equilibrium position and dynamics
of the O ion, allows two explanations of the largely
different T$_c$'s: (i) fine
details of the electronic structure at the Fermi level are very 
important, (ii) the very different degree of anharmonicity of the alkali ion 
is responsible for the differences in T$_c$.
We propose that the pressure dependence of T$_c$ can resolve 
these two scenarios. 
In scenario (i) a strong pressure dependence of T$_c$ with a negative slope
is expected following the trend accross the K-Rb-Cs series.
On the other hand the alkali ion dynamics, determined primarily
by its ionic radius, is not  
sensitive to small changes of the volume
and so no dramatic pressure dependence of $T_c$ is be expected if 
scenario (ii) applies.

J. K. was supported by DOE grant DE-FG03-01ER45876, while W. E. P. was
supported by NSF Grant DMR-0421810.


\begin{thebibliography}{10}

\bibitem{Ramirez}A. P. Ramirez,
 Annu. Rev. Mater. Sci.  {\bf 24}, 453 (1994).
\bibitem{Canals}B. Canals and C. Lacroix,
 Phys. Rev. B {\bf 61}, 1149 (2000).
\bibitem{Yonezawa} S. Yonezawa, Y. Muraoka, Y. Matsushita,
and Z. Hiroi, J. Phys. Condens. Matter {\bf 16}, L9 (2004).
\bibitem{Hiroi} Z. Hiroi, S. Yonezawa and Y. Muraoka, J. Phys. Soc. Japan
{\bf 73}, 1651 (2004).
\bibitem{rbos2o6} S.Yonezawa, Y.Muraoka, Y.Matsushita,
and Z.Hiroi, J. Phys. Soc. Jan. {\bf 73}, 819 (2004).
\bibitem{csos2o6}S. Yonezawa, Y. Muraoka, and Z. Hirio, cond-mat/0404220.
\bibitem{Hanawa} S.Hanawa, Y.Muraoka, T. Tayama, T.Sakakibara,
J.Yamaura, and Z.Hiroi, Phys. Rev. Let. {\bf 87},187001 (2001);
R. Jin, J. He, S. McCall, C. S. Alexander, F. Drymiotis, and
D. Mandrus, Phys. Rev. B {\bf 64} 180503 (2001).
\bibitem{bru04} M. Br\"uhwiler, S. M. Kazakov, N, D, Zhigadlo, J. Karpinski,
and B. Batlogg, Phys. Rev. B {\bf 70}, 020503 (2004).
\bibitem{Koepernik}K. Koepernik, and H. Eschrig,  
 Phys. Rev. B {\bf 59}, 1743 (1999);
 H.Eschrig, Optimized LCAO Method and 
the Electronic Structure of Extended Systems (Springer,Berlin,1989).
\bibitem{wien2k} P. Blaha, K. Schwarz, G. K. H. Madsen, D. Kvasnicka, 
  and J. Luitz, Wien2k,
  {\it An Augmented Plane Wave + Local Orbitals Program for Calculating 
  Crystal 
  Properties} (Karlheinz Schwarz, Techn. Universit\"at Wien, Wien, 2001), 
   ISBN 3-9501031-1-2.

\bibitem{perdew}J. P. Perdew and Y.Wang,  
 Phys. Rev. B
  {\bf 45}, 13244 (1992).

\bibitem{flatbands}S. Fujimoto, Phys. Rev. B {\bf 64}, 085102 (2001).
\bibitem{djs}D. J. Singh, P. Blaha, K. Schwarz, and I. I. Mazin,
  Phys. Rev. B {\bf 60}, 16359 (1999).
\bibitem{jan77}J. F. Janak, Phys. Rev. B {\bf 16}, 255 (1977);
S. H. Vosko and J. P. Perdew, Can. J. Phys. {\bf 53}, 1385 (1975)
\bibitem{sin02}D. J. Singh, P. Blaha, K. Schwarz, and J. O. Sofo, Phys. Rev. B
   {\bf 65}, 155109 (2002)
\end{thebibliography}
\end{document}